\documentclass[twocolumn,trackchanges]{aastex701}
\usepackage[table]{xcolor}
\usepackage{color, colortbl}
\usepackage{graphicx}
\usepackage{amsmath}
\usepackage{amssymb}
\usepackage{tabularx}
\usepackage{hhline}
\usepackage{makecell}
\usepackage{booktabs}
\usepackage{array}
\usepackage{acronym}
\usepackage{multirow}
\usepackage{booktabs}
\usepackage{xcolor}
\usepackage{graphicx}
\usepackage{tikz}
\usepackage{caption}

\definecolor{chmagenta}{rgb}{0.54, 0.17, 0.88}

\received{TBD}
\revised{TBD}
\accepted{TBD}

\begin{document}

\title{Twin Peaks: Resolving Features in the Binary Black Hole Mass Function with COSMIC-METISSE}

\author[0009-0000-8189-9887]{Duncan B. Maclean}
\affiliation{Department of Physics and Astronomy, University of North Carolina at Chapel Hill, 120 E. Cameron Ave, Chapel Hill, NC, 27514, USA}
\email{dmaclean@unc.edu}

\author[0000-0002-1135-984X]{Poojan Agrawal}
\affiliation{Institute of Astronomy, KU Leuven, Celestijnenlaan 200D, B-3001 Leuven, Belgium}
\email{poojan.agrawal@kuleuven.be}

\author[0000-0001-5228-6598]{Katelyn Breivik}
\affiliation{McWilliams Center for Cosmology and Astrophysics, Department of Physics,
Carnegie Mellon University, Pittsburgh, PA 15213, USA}
\email{kbreivik@andrew.cmu.edu}

\author[0000-0002-8304-0109]{Alexandra G. Guerrero}
\affiliation{Department of Physics, The University of Chicago, 5640 South Ellis Avenue, Chicago, Illinois 60637, USA}
\email{aghanselman@uchicago.edu}

\author[0000-0002-0147-0835]{Michael Zevin}
\affiliation{The Adler Planetarium, 1300 South DuSable Lake Shore Drive, Chicago, IL, 60605, USA}
\affiliation{Center for Interdisciplinary Exploration and Research in Astrophysics (CIERA), Northwestern University, 1800 Sherman Avenue, Evanston, IL, 60201, USA}
\email{mzevin@adlerplanetarium.org}

\author[0000-0002-6718-9472]{Mathieu Renzo}
\affiliation{Steward Observatory, Department of Astronomy, University of Arizona, 933 N. Cherry Ave., Tucson, AZ 85721, USA}
\email{mrenzo@arizona.edu}

\author[0000-0003-4175-8881]{Carl L. Rodriguez}
\affiliation{Department of Physics and Astronomy, University of North Carolina at Chapel Hill, 120 E. Cameron Ave, Chapel Hill, NC, 27514, USA}
\email{carl.rodriguez@unc.edu}

\begin{abstract}
Gravitational waves from inspiraling binary black holes (BBHs) provide insights into the lives and deaths of massive stars. Population synthesis allows us to model these binaries through isolated binary evolution, but its predictive power is limited by difficulties in varying the stellar models and their associated uncertainties. We present a new grid of stellar tracks computed with the open-source stellar evolution code \texttt{MESA}, spanning metallicities $10^{-3} \le Z/Z_{\odot} \le 7$. We vary two stellar physics parameters: wind-driven mass loss and the convective boundary mixing (CBM) mechanism. We pair these models with the Method of Interpolation for Single Stellar Evolution (\texttt{METISSE}) and binary population synthesis code \texttt{COSMIC} to obtain synthetic populations of merging BBHs in the local Universe. We find a maximum in the primary mass spectrum near $10M_\odot$ which in most model variations is composed of two sub-populations at $\approx8M_{\odot}$ and $\approx13 M_\odot$, with the higher-mass population dominated by BBHs whose progenitors underwent a mass ratio reversal (MRR). This population also suggests an anticorrelation between higher primary masses and mass ratio, as BBHs with $m_1\gtrapprox10M_\odot$ preferentially undergo MRR and prefer a final mass ratio of $q\approx0.7$. However, the location and relative strength of these two sub-populations is sensitive to our assumed stellar physics: varying both the wind and CBM treatments can merge the MRR and non-MRR populations into a single peak near $9M_\odot$. Variations in our stellar tracks, especially CBM, lead to a factor of $\approx6$ difference in the rate, primarily due to modulation of the common envelope formation channel.
\end{abstract}


\section{Introduction}\label{Intro}

In the last decade, gravitational-wave (GW) instruments such as Advanced LIGO, Virgo, and KAGRA (LVK) have enabled the detailed study of compact binary objects \citep{Abbott_2015}. The GW signals emitted by compact binary mergers contain information about the masses and spins of their binary components, which provide clues about their stellar progenitors. The most recent Gravitational Wave Transient Catalog (GWTC-5) has brought the number of confident GW detections to over 250, and several robust features of the observable binary black hole (BBH) population have emerged \citep{abac2025a_gwtc, abac2025b_gwtc, ligo2026a_gwtc, ligo2026b_gwtc}.

Population models of the LVK data have identified a global peak in the primary BH mass spectrum at $m_1\sim 10 M_{\odot}$, with a knee-like feature at $m_1\sim 35M_{\odot}$. These features bound a power law-like mass spectrum which steepens above $\sim 35 M_{\odot}$ \citep{abbott2023population, Callister2024-qz, abac2025b_gwtc, ligo2026b_gwtc}. The component mass ratio ($q \equiv m_2/m_1$) of merging BBHs displays a feature at $q\approx0.7$ which is preferentially identified with BBHs near the $10 M_{\odot}$ peak \citep{abac2025b_gwtc}. Higher-mass BHs, meanwhile, are found to preferentially merge with more equal-mass companions, potentially evidencing dynamical assembly in dense stellar clusters \citep{abac2025b_gwtc}. Most proposed channels for BBH formation rely on either isolated binary evolution or dynamical dynamical assembly, \citep{Mapelli_2021, Mandel_2022, ray2026astrophysicaloriginbinaryblack}, but significant deficiencies remain in our understanding of how these massive stellar progenitors evolve \citep{Mapelli_2020b, Zevin_2021,Cheng_2023}.

Stars massive enough to form BHs experience strong wind-driven mass loss throughout their lives, affecting their interior structure, terminal mass, and the mass ejected during the supernova explosion \citep{Renzo2017}. On the main-sequence, these winds are predominantly line-driven and metallicity-dependent \citep{Lucy1970-jr}, while evolved stars experience multiple mass-loss processes associated with dust and pulsational episodes \citep{romagnolo2026stellarwindsatlasi, Yoon_2010}. Historically, all of these regimes have been poorly constrained, but this situation has improved in recent years \citep{romagnolo2026stellarwindsatlasi}.

The physics of the stellar interior, such as supernova explosion mechanisms and energy transport, present their own challenges. Specifically, convective energy transport is deeply relevant to the formation of the stellar core, and uncertainties in this process propagate to the final remnant of a star. Convection is an inherently 3-dimensional process, and 1D approximations such as mixing length theory (MLT) alone are insufficient to model the various processes that transfer heat and elements through the star. Thus, prescriptions for convective boundary mixing (CBM) are required to treat mixing in these regions \citep{herwig2000, Joyce2023}.  CBM refers to a group of processes which allow for the transfer of chemical species, entropy, and heat between the convection-dominated and radiation-dominated zones \citep{Anders_2022}. In massive stars with convective cores, accurate modeling of CBM is critical to determining the extent and properties of the stellar core.

Uncertainties in these intrinsic processes propagate into binary evolution, significantly affecting models of BBH formation. Stellar mass loss constrains the maximum mass of a given star's remnant while also sapping angular momentum from the binary orbit. Convective boundary mixing, meanwhile, directly modifies the star's core mass and radius, affecting the likelihood of Roche lobe overflow (RLOF) and mass transfer \citep{Agrawal_2023}. Mass transfer, in turn, can harden a binary system, cause it to merge, or even cause a mass ratio reversal (MRR) where the star that is initially more massive forms the lower mass compact object \citep{Zevin_2022, Broekgaarden_2022, Smith_2026, Chen+2026:2026arXiv260626262C}. To fully understand the range of possible outcomes of binary stellar evolution, we must first understand how the uncertainties in massive star evolution propagate to stellar populations at large.

Exploring the isolated binary evolution channel for merging BBHs---our focus in this study---requires stellar and binary population synthesis tools capable of rapidly modeling a progenitor binary system throughout its entire evolution with varying input physics. Rapid population synthesis codes such as \texttt{binary\_c} \citep{Izzard2004}, \texttt{Startrack} \citep{Belczynski_2008}, \texttt{SeBa} \citep{Toonen2012}, \texttt{COSMIC} \citep{Breivik_2020} and \texttt{COMPAS} \citep{COMPASTeam:2021tbl} integrate the Single-Star Evolution (\texttt{SSE}) fitting formulae to obtain the properties of the stellar components \citep{Tout1997, Pols_1998, Hurley_2000}. Significant progress has been made in the modeling of massive stars since these tracks were introduced, and recent additions to the field such as \texttt{SEVN} \citep{Spera2015, Iorio2023}, \texttt{METISSE} \citep{Agrawal_2020, Agrawal_2023, Agrawal_2025} and \texttt{ComBinE} \citep{Kruckow_2018} instead interpolate directly between pre-calculated single stellar tracks, while \texttt{POSYDON} \citep{Fragos_2023, Andrews_2025} interpolates between binary stellar tracks. Interpolation provides an opportunity to directly study how variations in stellar evolution propagate to binary populations and, ultimately, compact binary mergers.

In this study, we use an interpolation-driven population synthesis pipeline to investigate how uncertainties in massive stellar evolution affect the rates and demographics of BBH mergers. In Section \ref{Methods}, we detail our tool chain, which includes the simulation of 1D stellar models with \texttt{MESA}. We then incorporate these models into the rapid population synthesis code \texttt{COSMIC} \citep{Breivik_2020} using the Method of Interpolation for Single Star Evolution \citep[\texttt{METISSE},][]{Agrawal_2025}. Section \ref{S3_BBH_demography} details our findings for BBH merger rates, the BBH primary mass distribution, and the component mass ratio for each physics variation. Finally, in Section \ref{Disc_SubPops_10Msun} we interpret the two major features present in our BBH mass spectra and identify the dominant formation pathway for both sub-populations. We detail the roles of MRR, stable mass transfer (SMT), and common envelope (CE) evolution in the process, and their influence on the shape of the $10M_{\odot}$ peak(s).

\section{Method}\label{Methods}

We produce stellar tracks with the open-source stellar evolution code \texttt{MESA}, release \texttt{r24.08.1} \citep{Paxton2011, Paxton2013, Paxton2015, Paxton2018, Paxton2019, Jermyn2023}. \texttt{MESA} provides several thread-safe software modules which cooperate to evolve a stellar model. The core \texttt{star} module solves the equations of stellar structure on a coupled, fully Lagrangian mesh of radial ``zones." The \texttt{eos} module obtains physical quantities such as temperature, density, and pressure at the faces and center of each zone \citep{Rogers2002, Saumon1995, Irwin2004, Timmes2000, Potekhin2010, Jermyn2021}. Opacity information is provided by the \texttt{kap} module \citep{Iglesias1993, Iglesias1996, Ferguson2005, Poutanen2017, Cassisi2007, Blouin2020}. Nuclear burning and the associated reaction rates are implemented through the \texttt{net} module \citep{Cyburt2010, Angulo1999, Fuller1985, Oda1994, Langanke2000, Chugunov2007, Itoh1996}, while \texttt{mlt} resolves convective energy transport and mixing. \texttt{MESA} is highly configurable, providing hooks for users to replace functionality from the above modules with custom physics.

\subsection{Adopted Physics \& Grid Variations}\label{Method_Physics}

Our choice of input settings is informed by currently available tracks. We use the input physics of \texttt{POSYDON} \citep{Fragos_2023, Andrews_2025} as the basis for our variations. We adopt a simple $T(\tau)$ atmosphere using the Eddington grey approximation. For convective mixing, we use the \cite{mihalas78} mixing length theory prescription, with $\alpha_{\rm MLT} = 1.93$. To assist convergence, we enable \texttt{okay\_to\_reduce\_gradT\_excess}, which enhances convective energy transport in regions with a superadiabatic temperature gradient \citep{Paxton2013}. We also set a global mixing floor, $D_\text{min}=10^{-2} \text{cm}^2 \text{s}^{-1}$ to smooth local chemical gradients\footnote{For a detailed discussion of $D_\text{min}$, we refer the reader to \cite{Farag_22}, Section 3.1}. For semiconvective mixing, we adopt $\alpha_{\rm semi}=10^{-4}$. We do not model thermohaline mixing or element diffusion. We discuss our implementations of convective boundary mixing in Section~\ref{Methods_CBM}.

Our \texttt{MESA} tracks include $85$ mass points. These are stepped from $0.1\leq m/M_\odot \leq 1.0$ in $0.1M_\odot$ intervals, log-spaced from $1.0 \leq m/M_{\odot} \leq 150$, and stepped again in $25 M_\odot$ intervals from $150.0 \leq m/M_\odot \leq 300.0$. We evolve an additional $42$ naked helium stars ($X=0$) at $0.1 \leq m_{\rm He}/M_{\odot} \leq 150$. These permit modeling of stellar cores that become stripped through binary evolution. Our tracks span 28 metallicities between $10^{-3} \leq Z/Z_{\odot} \leq 7$, although for this study we only employ tracks up to $2 \times Z_\odot$. We evolve our stellar models until central carbon depletion, less than $\sim 1$ thermal timescale before the supernova. For low-mass stars which do not ignite carbon, we terminate the evolution at the time of envelope loss, around the late asymptotic giant branch (AGB). We evolved four variations of these \texttt{MESA} tracks for a grand total of 14,224 stellar models. These models significantly exceed the parameter space covered by the well-utilized \cite{Pols_1998} stellar tracks, which include $m=0.5-50 M_\odot$ and $0.01 \leq Z/Z_\odot \leq 1.5$ (for their $Z_\odot = 0.02$).

In Table~\ref{Table1}, we provide an overview of the physics prescriptions in each of our model variations. Our \textsc{standard} differs from \cite{Fragos_2023} through the addition of a wind prescription for luminous blue variable (LBV) stars, identical to that used in \cite{Andrews_2025}. From this baseline we vary two key stellar evolution parameters: the wind mass loss formula for massive stars (``winds''), and the convective boundary mixing (``CBM'') prescription at the core-envelope boundary. Wind mass loss can strip a star's envelope as it ages, setting an effective upper limit on the pre-supernova core mass, which therefore limits the size of a compact remnant. CBM, meanwhile, models the exchange of chemical species, heat, and entropy across the boundaries between convective- and radiation-dominated regions. This has implications both for the duration of main-sequence burning and for the core mass at the time of supernova. A detailed comparison between our models, which interpolate single-star tracks, and \texttt{POSYDON} models, which interpolate between binary star tracks, is currently in preparation \citep{guerrero2026:inprep}.

In massive stellar binaries, these phenomena influence whether a given star will overflow its Roche lobe, and whether the ensuing mass transfer will be dynamically stable. We discuss the models applied for winds and CBM in detail in Sections \ref{Methods_Winds} and \ref{Methods_CBM} below, respectively.

\begin{table*}[htbp]
  \centering
    \begin{tabular}{lccc}
        MESA Grid & Hot wind & Cool wind & CBM \\
        \hline
        \textsc{standard} & \cite{Vink2001-af} & \cite{deJager88} & \cite{herwig2000} \\
        \textsc{new winds} & \cite{Krticka2025-rn} & \cite{Decin2023-je} & \cite{herwig2000} \\
        \textsc{new CBM} & \cite{Vink2001-af} & \cite{deJager88} & \cite{Johnston2024-bf} \\
        \textsc{new winds \& CBM} & \cite{Krticka2025-rn} & \cite{Decin2023-je} & \cite{Johnston2024-bf} \\
        \hline
    \end{tabular}
    \caption{Physics variations in our \texttt{MESA} model grids. All other input settings are identical to those used in \cite{Fragos_2023} except for our addition of clumpy, luminous blue variable (LBV) winds for hot, very massive stars which exceed the Eddington limit \citep{Belczynski_2008}. This change makes our \textsc{standard} model most comparable to the stellar tracks used in \texttt{POSYDON V2} \citep{Andrews_2025}.}
    \label{Table1}
\end{table*}

\subsubsection{Wind Mass Loss Rates for Massive Stars}\label{Methods_Winds}

Both of our wind choices re-implement \texttt{MESA}'s ``Dutch'' wind subroutine. This provides both ``hot'' ($T_{\rm eff} \geq 11,000$ $K$) and ``cool'' ($T_{\rm eff} \leq 10,000$ $K$) mass loss schemes and smoothly blends between them when $10,000 < T_{\rm eff}/K < 11,000$. Our \textsc{standard} model adopts the \cite{Vink2001-af} winds for hot (OB) stars, and the \cite{deJager88} prescription for red supergiants. Our ``new'' wind model, meanwhile, adopts two contemporary, observationally-calibrated wind prescriptions. For OB stars we adopt the fit provided by \cite{Krticka2025-rn}, while low-temperature winds are provided by the CO(2-1) line emission-derived fit from \cite{Decin2023-je}. To prevent undesired extrapolation, we clip these winds to within the parameter limits of their respective samples. Specifically, we limit the $T_\text{eff}$ contribution for \cite{Krticka2025-rn} winds to $\leq 45,000 \space \rm K$ and the metallicity contribution to $10^{-2} \leq Z/Z_\odot \leq 1$. For \cite{Decin2023-je} red supergiant (RSG) winds, we limit the luminosity contribution to $L \leq 5.56 \times 10^5 \space L_\odot$, and the initial mass contribution to $\leq 30 \space M_\odot$. 

All models employ two additional wind prescriptions which are not varied here. We implement the \texttt{StarTrack} wind prescription described in \cite{Belczynski_2010} for very massive luminous blue variables (LBVs) which cross the empirical Humphreys-Davidson limit \citep{Humphreys_1994}:

\begin{equation}
    \dot{M}_\text{LBV} = 1.0\times10^{-4} \space
    [M_\odot \space \text{yr}^{-1}]
\end{equation}

For evolved, hydrogen-poor stars ($X_\textrm{surf}<0.4$), we switch to the optically thick, Wolf-Rayet-like wind of \cite{NugisLamers_2000}, which parameterize the mass loss by the stellar luminosity, helium, and metal content:

\begin{equation}
    \dot{M}_\text{WR} = 1.0\times10^{-11}
    (L/L_\odot)^{1.29} \space Y^{1.7} \space Z^{0.5}
    \space[M_\odot \space \text{yr}^{-1}]
\end{equation}

\noindent where $L$ is the stellar luminosity, $Y$ is the total helium fraction, and $Z$ is the metallicity.

Our wind choices are motivated by the discrepancy between theoretically-calculated OB wind mass loss rates \citep{Vink2001-af} and estimates derived from contemporary observations \citep{Krticka2023-fy, Krticka2025-rn}. For hot, main-sequence stars, radiation-driven winds can strip the star's envelope and reduce the maximum radius it attains as a giant. Similarly, strong low-temperature winds such as those predicted by \cite{deJager88} efficiently strip the hydrogen-rich envelopes of red supergiants at high metallicity, causing these stars to evolve into blue or yellow supergiants before perishing \citep{Georgy_2012}.

\begin{figure*}[p!]
    \centering
    \includegraphics[width=\textwidth]
    {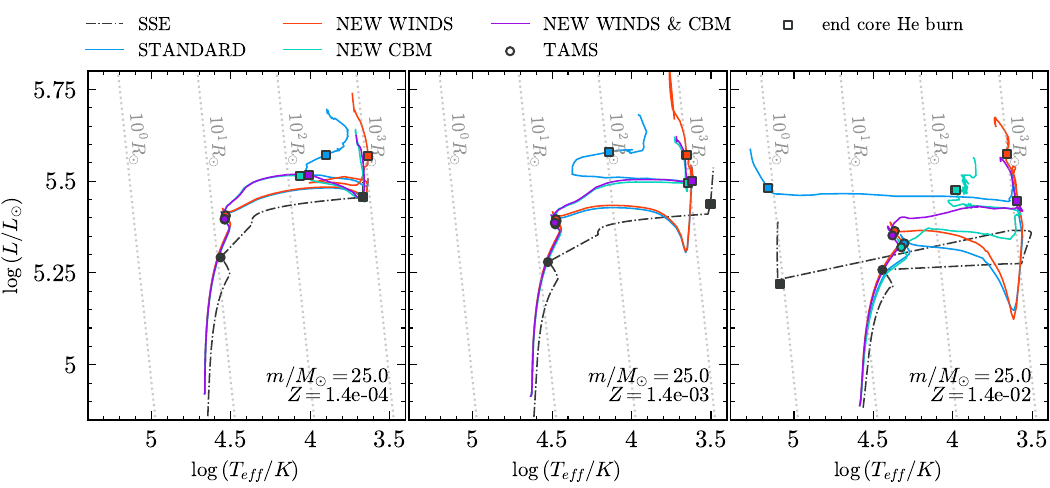}
    \caption{Stellar tracks interpolated with \texttt{COSMIC-METISSE} for a 25~$M_\odot$ star, obtained from tracks using each of our stellar grid variations. The dashed line shows a 25~$M_\odot$ track produced by SSE \citep{Pols_1998, Hurley_2000}. Post-main-sequence behavior differs significantly in each model, and the divergence between tracks correlates with increasing birth metallicity. At solar-like ($Z=0.014$, right panel) metallicity, \textsc{new winds} predict incomplete envelope stripping and red supergiant evolution \citep{Decin2023-je, Krticka2025-rn}.  Both the maximum stellar radius, which we show in Figure~\ref{Fig_Rmax}, and the evolutionary phase at which this radius is attained vary depending on the metallicity, winds, and CBM prescription.}
    \label{Fig_HR}
    
    \centering
    \includegraphics[width=1\textwidth]
    {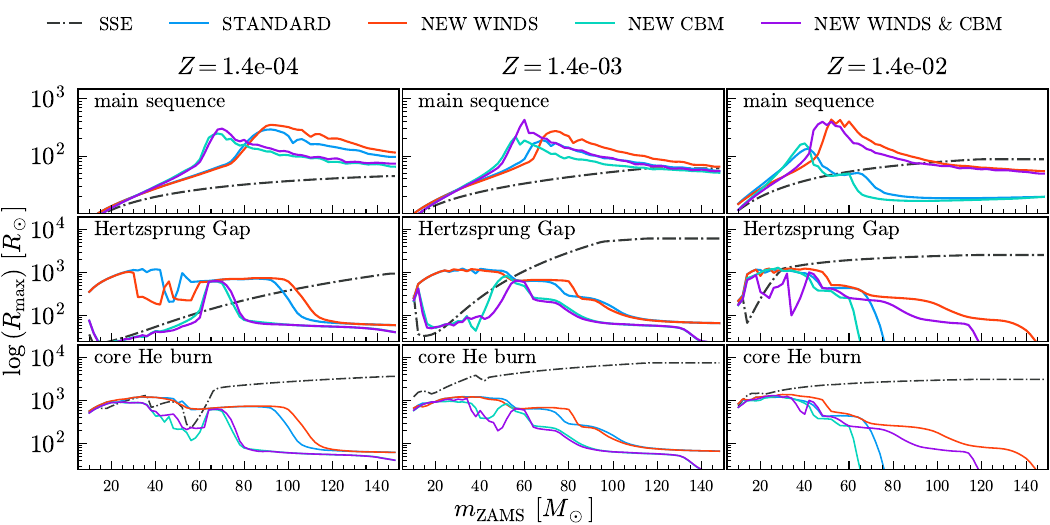}
    \caption{The maximum radii of stellar tracks for $10.0 \leq m_\text{ZAMS}/M_\odot \leq 150.0$ interpolated with \texttt{COSMIC-METISSE}. Maximum radii calculated with \texttt{METISSE} diverge sharply from the \texttt{SSE} fitting formulae, which are extrapolated at masses greater than 50 $M_\odot$. The radial growth of stars in each model shows strong mass and metallicity dependence, but we do observe several trends. At low metallicity, the CBM implementation is the dominant factor in determining the star's maximum radius. Stars evolved with convective penetration (\textsc{new CBM}) show a steeper radial growth in mass space on the main-sequence (top row), but this relationship inverts at very high masses. Models using convective penetration (\textsc{new CBM}) expand slowly throughout the Hertzsprung gap and attain their maximum radii during core helium burning. At solar metallicity (right column), the maximum radius is chiefly set by wind mass loss, which efficiently shrinks very massive stars while they are still on the main-sequence.}
    \label{Fig_Rmax}
\end{figure*}

\subsubsection{Convective Boundary Mixing}\label{Methods_CBM}

One-dimensional approximations of convection such as MLT do not account for mixing across convective boundaries, such as the edge of a massive, convective core which borders a radiative envelope. Additional treatment of these regions, which we refer to as ``convective boundary mixing,'' or CBM, is therefore required.

MLT aphysically assumes that a parcel of convective material which reaches a convective boundary has zero kinetic energy; in fact, such a parcel is likely to pass this boundary and diffuse its contents into the radiation-dominated zone. The extent and efficiency of such mixing can be estimated through several processes.

The most well-explored theory of CBM, referred to as ``convective overshoot'' or simply ``overshooting,'' permits the exchange of chemical species over convective boundaries. When applied to a massive stellar core, this prolongs the main-sequence lifetime and circulates the material neighboring the convective core. More efficient overshooting, therefore, leads a star to develop a more massive core.

Our \textsc{standard} and \textsc{new winds} models adopt a diffusive overshoot prescription which extends mixing into the radiative zone via the exponential-decay formalism in \cite{herwig2000} (see their equation 2). We parameterize the overshoot efficiency parameter, $f_{\rm ov}$, by ZAMS mass in the manner following \cite{Choi_2016, Fragos_2023, Andrews_2025}:

\begin{equation}\label{Eq_f_ov}
        f_{\rm ov} = \begin{cases} 
          0.016 & m \le 4M_{\odot} \\
          0.016 + (0.0415-0.016) \times \gamma & 4 < m/M_{\odot} < 8 \\
          0.0415 & m \ge 8 M_{\odot}
        \end{cases}
\end{equation}

\noindent where
\begin{equation*}
        \gamma = \frac{1}{2}\Big[ 1 - \cos{\big(\pi(\frac{m}{4}-1)\big) \Big].}
\end{equation*}

Our \textsc{new CBM} model adopts the extended convective penetration mixing scheme of \cite{Anders2022-yz}. In contrast to exponential overshooting, which alters chemical gradients adjacent to the convective boundary, convective penetration models the transfer of heat and entropy within a fully-mixed ``penetrative zone.'' This zone extends from the star's convective core into the neighboring, radiation-dominated layer, transporting chemical species, heat, and entropy. The radial extent of the penetrative zone is determined by balancing the diffusive and buoyant work done in the convective and penetrative zones against the negative buoyant work done by overdense material in the penetrative zone. The fully-mixed penetrative zone terminates with a thin, partially-mixed overshoot zone, which blends smoothly into the radiative layer \citep{Johnston2024-bf}.

In Figure~\ref{Fig_HR} we visualize the evolutionary effects of our stellar physics variations in a star with $M_{\rm ZAMS}=25 M_{\odot}$. At solar metallicity (right-hand panel), the \textsc{standard} model using the winds of \cite{Vink2001-af} and \cite{deJager88} loses its envelope through winds alone and evolves into a hot, Wolf-Rayet-like star. The \textsc{new winds} model \citep{Krticka2025-rn, Decin2023-je} experiences incomplete envelope stripping, evolving instead into a red supergiant. At low metallicity, the \textsc{standard} model undergoes a blue loop phase, while the \textsc{new winds} model remains a red supergiant. Additionally, convective penetration \citep{Johnston2024-bf}, or, \textsc{new CBM}, shifts the location of the terminal main-sequence hook and the duration of the star's main-sequence phase.

In Figure~\ref{Fig_Rmax}, we show the maximum stellar radius attained as a function of ZAMS mass, both during and after the main-sequence. At low metallicity (left panel), the CBM prescription is the primary controller of a star's maximum radius; meanwhile, at high metallicity, efficient wind-driven mass loss becomes the primary constraint on radius. Stars evolved with convective penetration (\textsc{new CBM}) attain larger radii while on the main sequence until $\approx 50 M_\odot$, when this relationship inverts. At solar metallicity (right panel), winds become the primary constraint on radius.


\subsection{Stellar Track Interpolation with \texttt{METISSE}}\label{Methods_METISSE}

We use \texttt{METISSE} \citep{Agrawal_2025} to replace the single star component of our population synthesis pipeline. \texttt{METISSE} ingests pre-computed stellar evolutionary tracks and interpolates between them to produce a new track with a user-supplied ZAMS mass and age. For use with \texttt{METISSE}, we have converted our MESA-evolved stellar tracks into the Equivalent Evolutionary Point (\texttt{eep}) format using the code package \texttt{iso} \citep{Dotter_2016}. The combination of \texttt{COSMIC} with \texttt{METISSE} as a stellar evolution backend, (henceforth \texttt{COSMIC-METISSE}), allows us to perform rapid population synthesis while exploring the the impacts of our model variations. This also permits us to model stars outside the metallicity bounds in the \texttt{SSE} formulae, which are not supported outside $5\times10^{-3} \le Z/Z_{\odot} \le 1.5$ (for \texttt{SSE}'s assumed solar metallicity of $Z_{\odot} = 0.02$). Our \texttt{MESA} input files and the processed \texttt{eep} history files are publicly available \citep{S3_XXQBXM_2026}.

\begin{figure*}[t]
    \centering
    \includegraphics{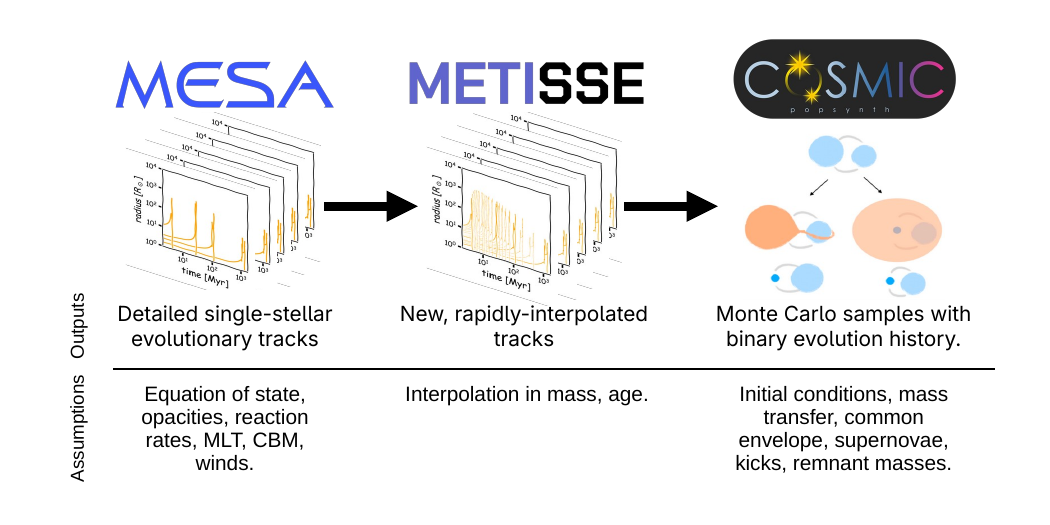}
    \caption{A schematic representation of our population synthesis pipeline and the uncertainties associated with each step. Stellar tracks evolved with \texttt{MESA} (left) provide us with stellar parameters for a given ZAMS mass, time, and metallicity. \texttt{METISSE} (middle) interpolates between these tracks to enable continuous sampling in mass. \texttt{COSMIC} (right) then samples the full population of initial masses, orbital periods, and eccentricities before evolving binaries using these tracks.}
    \label{Fig_Schematic}
\end{figure*}

\subsection{Rapid Population Synthesis with \texttt{COSMIC-METISSE}}\label{Method_COSMIC}

We use the \texttt{COSMIC-METISSE} framework as implemented in version \texttt{v4.0.1} of \texttt{COSMIC} to perform binary population synthesis simulations. \texttt{COSMIC} samples the initial conditions of binary systems and then evolves them with a rapid binary evolution algorithm based on the high-performance Binary Stellar Evolution (\texttt{BSE}) code \citep{Hurley_2002}. \texttt{COSMIC} contains several upgrades to the treatment of massive star evolution, compact object formation and binary interactions; for a list of these upgrades and subsequent codebase history, see \citet{Breivik_2020}\footnote{The \href{https://cosmic-popsynth.github.io/}{\texttt{COSMIC} documentation} contains a complete list of changes.}.

\texttt{COSMIC} provides various settings controlling the binary interactions between stars in \texttt{BSE}. We show a sample of these settings in Table~\ref{Tab_BinaryParams}. When a star overflows its Roche lobe, we follow \cite{Hurley_2002} and limit the mass accreted by the companion based on an efficiency factor, $\beta_{acc}$. This factor is derived based on the accretor's thermal timescale:

\begin{equation}
    \beta_\textrm{acc} = \rm min\left(
    10 \times \frac
    {\tau_{\dot{M}}}
    {\tau_{\textrm{KH}}}
    , 1\right)  
    \label{Eq_beta_acc}
\end{equation}

\noindent where $\tau_{\dot{M}}$ is the accretor's accretion timescale and $\tau_{\textrm{KH}}$ is its Kelvin-Helmholtz timescale. For giant stars with diffuse envelopes, we apply no limit, as their thermal timescales are comparatively short \citep{Hurley_2002}.

For the CE phase, we employ the $\alpha-\lambda$ formalism \citep{Webbink1984, deKool1990}. The efficiency factor, $\alpha_\text{CE}$, sets the efficiency with which orbital energy expels the envelope, while the factor $\lambda_\text{CE}$ accounts for the envelope binding energy. We adopt $\alpha_\text{CE}=1$. For our reference \texttt{SSE} model, we take $\lambda_\text{CE}$ from the stellar density profile, following \cite{Claeys_2014}. The \texttt{COSMIC-METISSE} model variations instead directly integrate the gravitational binding energy of the interpolated stellar track's envelope:

\begin{equation}
    E_\text{bind} = \int_{R_{c}}^R \left( E_k(r) - \frac{Gm(r)}{r} \right)dr
\end{equation}

\noindent where $E_k(r)$ and $m(r)$ are the kinetic energy and mass at $r$, $R_c$ is the outermost radius of the stellar core, $G$ is the gravitational constant, and $R$ is the stellar radius.

For BH formation, we adopt the ``delayed" remnant prescription of \cite{Fryer_2012}, which models post-supernova fallback onto the nascent remnant. Very high mass stars which undergo pulsational pair-instability supernovae are shrunk or destroyed following \cite{Renzo2022, Hendriks2023}. Following \cite{Belczynski_2008}, we assume a ``pessimistic" CE scenario, in which donor stars which lack a core-envelope boundary (e.g.~main-sequence or Hertzsprung gap) are assumed to always merge during a CE phase. All other binary settings follow the default \texttt{Params.ini} file for \texttt{COSMIC v.4.0.1}. 

\startlongtable
\begin{deluxetable*}{lcl}
    \tablewidth{\linewidth}
    \tabletypesize{\small}
    \tablecaption{A sample of our adopted \texttt{COSMIC} binary evolution parameters that play a significant role in the formation of BHs. Our complete inifiles are publicly available \citep{duncan_b_maclean_2026_20881972}.}
    
    \tablehead{
        \colhead{\texttt{COSMIC} setting} & \colhead{adopted}  & \colhead{description}
        }
        
    \startdata
        \texttt{alpha1} & 1.0 & \parbox[t]{5.5in}
        {
            During CE, orbital energy is used to expel the envelope with
            perfect efficiency \citep{Pacznski76}.
        } \\
        \texttt{lambdaf} & 1.0 & \parbox[t]{5.5in}
        {
            During CE, the envelope binding energy is interpolated from
            stellar tracks using \texttt{METISSE}. In the \texttt{SSE} reference model, which does not have a pre-computed binding energy, we use 
            the fit of \cite{Claeys_2014} instead.
        } \\
        \texttt{cemergeflag} & 0 & \parbox[t]{5.5in}
        {
            CEs with a donor lacking a core-envelope boundary are not forced to merge,
            but we remove these systems in post-processing \citep{Belczynski_2008, Dominik_2012}.
        } \\
        \texttt{pisn} & -4 & \parbox[t]{5.5in}
        {
            Pulsational pair-instability supernovae reduce the compact remnant mass based on a fit derived
            from the carbon-oxygen core mass and metallicity \citep{Renzo2022, Hendriks2023}.
        } \\
        \texttt{remnantflag} & 4 & \parbox[t]{5.5in}
        {
            The ``delayed'' prescription of \cite{Fryer_2012}, which produces no mass gap between
            neutron stars and BHs.
        } \\
        \texttt{kickflag} & 5 & \parbox[t]{5.5in}
        {
            BH natal kicks are drawn from a log-normal distribution with $\mu=5.67$ and $\sigma=0.59$ \citep{Disberg_2025}.
        } \\
        \texttt{bhflag} & 1 & \parbox[t]{5.5in}
        {
            BH natal kicks are reduced by the fraction of the mass that falls back onto the proto-compact object where the fallback mass is determined by \citet{Fryer_2012}.
        } \\
        \texttt{qcflag} & 5 & \parbox[t]{5.5in}
        {
            Mass transfer is considered ``stable'' if $m_\text{donor}/m_\text{accretor} < q_{\rm crit}$, where $q_{\rm crit}$ is a function of evolutionary stage. Mass transfer from stripped stars is assumed to always be stable \citep{Neijssel_2019}. 
        } \\
        \texttt{acc\_lim} & -1 & \parbox[t]{5.5in}
        {
            The mass accreted during Roche lobe overflow is limited to $10$ times the thermal timescale-derived limit of the accretor, as defined in \cite{Hurley_2002} (though see \S\ref{Disc_SMT_early} for variations).
        }
    \enddata
    
    \label{Tab_BinaryParams}
\end{deluxetable*}

Binaries are evolved this way in batches until the stellar population is converged such that further binaries do not modify the population characteristics. We employ \texttt{COSMIC}'s built-in independent sampler, with an initial mass function from $0.1\leq m_\text{ZAMS} \leq 150$ following \cite{kroupa2001} and a flat $q$ distribution. Orbital periods and eccentricities are drawn from a power law following \cite{Sana_2012}. We assume a fixed binary fraction of $0.7$ and discard single systems, but we preserve the total mass formed in order to obtain population realizations (see \cite{Breivik_2020} \S 2.2). We provide a schematic cartoon of our population synthesis pipeline in Figure~\ref{Fig_Schematic}, showing the progression from stellar tracks evolved with \texttt{MESA} to population realizations generated with \texttt{COSMIC-METISSE}.

\subsection{Cosmic Star Formation History}\label{Methods_SFH}
Predicting the properties of GW transients requires that we account for the cosmic star formation history (SFH). This entails three components, namely the redshift-dependent star formation rate density (SFD), the redshift-dependent metallicity dispersion, and a cosmological model. We then convolve these to obtain the rate density of stellar mass formed at a given metallicity at a given comoving time. We adopt the ``moderate'' model of \cite{Chruslinska19}, which provides the SFD and metallicity dispersion $\Psi(Z,z)$ from $0 \le z \le 10$.
For all cosmological calculations we assume a flat, $\Lambda \text{CDM}$ cosmology with $H_0=67.74$ $\text{km\,s}^{-1}\text{Mpc}^{-1}$, and $\Omega_{m}=0.3075$ \citep{Planck15CosParams}.

\subsection{GW Event Rates}\label{Methods_Rates}

Having obtained a Monte Carlo sample of BBH mergers and the metallicity-dependent cosmic star formation history, we estimate BBH merger rates in a manner following \cite{Dominik_2013}. For a given comoving time coordinate $t$, one must calculate the rest-frame merger rate $n_\textrm{rest}$:

\begin{equation}
    n_{\rm rest}(t_i) = \sum_k^N \mathcal{R}_{\rm form,k,i} (t_i - t_{\rm delay,k}) \space~ [\rm Gpc^{-3} \space yr^{-1} ],
    \label{Eq_n_rest}
\end{equation}
\noindent where
\begin{equation}
    \mathcal{R}_{\rm form,k,i} = \\
    \xi_k \times \\
    \frac{f_{\rm binary}}{M_{\rm sim}} \\
    \times \Psi(Z_k,z_i) \\
    \space~ [\rm Gpc^{-3} \space yr^{-1}]
    \label{Eq_R_form}
\end{equation}

\noindent and $\xi_k$ is the dimensionless ``weight'' of a given binary $k$ represented in our Monte Carlo sample, $\Psi(Z_k,z)$ is the metallicity-dependent star formation rate density for this binary at redshift $z_i$, $M_\text{sim}$ is the total stellar mass formed in our sample, and $f_{\rm binary}$ corrects for the fraction of stars formed in binaries (in our case, $f_{\rm binary}=0.7$).

To obtain merger rates we select $N_i=300$ equally-spaced time bins spanning $0\le z \le 10$. We then calculate the above for each system $k$ at the center of each comoving time bin $i$. Naturally, binaries that do not merge in a Hubble time, or that formed in a bin too late to permit a merger, are discarded. We then obtain a local, observable merger rate by summing the rate of events over the local ($z<0.2$) cosmological volume, per \cite{Dominik_2013}:

\begin{equation}\label{Eq_R_obs}
    \mathcal{R}_{\text{obs},k}(<z) = 4\pi \int_0^z \frac{n_{\text{rest},k} }{1 + z'} \frac{dV}{dz'}dz' \ [\text{yr}^{-1}]
\end{equation}

\noindent where $n_{\text{rest},k}$ is the result of Equation \eqref{Eq_n_rest}, $1/(1+z')$ is the correction between cosmic time at redshift $z$ and the observer time at redshift zero, and $dV/dz'$ is the differential comoving volume element.  We implement the above equations in our own rates integrator code, \texttt{MC-rates}, which provides merger rates and component masses for each binary in our Monte Carlo sample, forming a map between these traits and their initial conditions. These data enable us to study the impact of our stellar physics variations on individual binaries and on the general BBH population. The source code for \texttt{MC-rates} is publicly available as a python module. \citep{maclean_2026_20799423}.

\section{Local BBH Merger Rates and Demographics}\label{S3_BBH_demography}

\subsection{The Local Merger Rate}\label{S3_Rates}

\begin{figure}[htbp]
    \centering
    \includegraphics{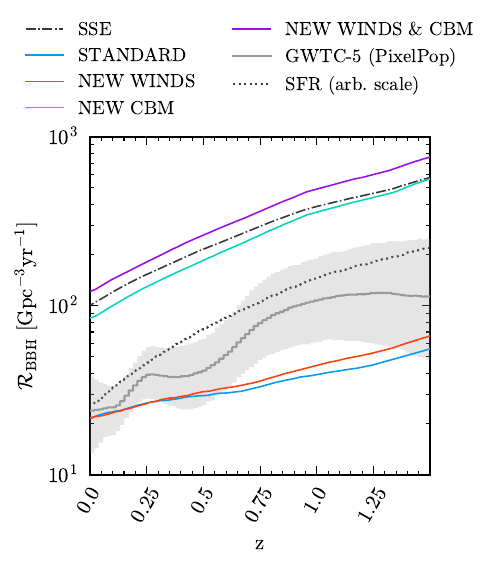}
    \caption{The volumetric BBH merger rate as a function of redshift. The LVK \textsc{PixelPop} posterior is given in grey, and the shaded region denotes the $90\%$ CI. Our \texttt{SSE} reference model is given by the dashed line. Our \textsc{standard} and \textsc{new winds} models  intersect the median LVK rate estimate at $z\lessapprox0.2$. The \textsc{new CBM}, \textsc{new winds \& CBM}, and \textsc{SSE}, models better trace the redshift-dependent star formation history (dot-dashed black line) \citep{Chruslinska19}, while the \textsc{standard} and \textsc{new winds} models display a shallow redshift evolution.}
    \label{Fig_rates}
\end{figure}

We report the estimated local ($z=0.2$), intrinsic BBH merger rate for each of our models in Table \ref{Tab_rates}. Our estimate for the local BBH merger rate is highly model-dependent, and no \texttt{COSMIC-METISSE} model recovers both the LVK local merger rate and the redshift evolution of the rate. Our \textsc{standard} and \textsc{new winds} both predict a BBH merger rate of $\approx25$ $\text{Gpc}^{-3}\text{yr}^{-1}$, while our models using convective penetrative mixing (\textsc{new CBM} and \textsc{new winds \& CBM}) yield substantially greater rates: $120$ and $168~\rm Gpc^{-3} yr^{-1}$, respectively. In Figure~\ref{Fig_rates} we show the volumetric BBH merger rate as a function of redshift out to $z = 1.5$, with the LVK posterior and the (arbitrarily scaled) star formation density shown for reference.

\begin{table}[htb]
  \centering
    \begin{tabular}{lc}
        \textbf{Model} & $\mathcal{R}_\text{BBH} \space |_{z=0.2}$ [$\text{Gpc}^{-3} \text{yr}^{-1}$] \\
        \hline
        {GWTC-5 (\textsc{PixelPop})} & $37.5^{+11.9}_{-9.1}$ \\
        \hline
        \textsc{standard} & 25.64 \\
        \textsc{new winds} & 25.62 \\
        \textsc{new CBM} & 119.83 \\
        \textsc{new winds \& CBM} & 168.25 \\
        \textsc{SSE} (reference) & 142.64 \\
        \hline
        \textit{Optimistic CE} & \\
        \textsc{standard} & 419.44 \\
        \textsc{new winds} & 388.92 \\
        \textsc{new CBM} & 399.37 \\
        \textsc{new winds \& CBM} & 381.13 \\
        \textsc{SSE} (reference) & 381.46 \\
        \hline
    \end{tabular}
    \caption{The estimated local ($z=0.2$) BBH merger rate calculated with \texttt{MC-Rates}. The higher merger rate obtained with \textsc{new CBM} is predominantly due to an abundance of systems which harden during CE episodes; comparable systems in the \textsc{standard} and \textsc{new winds} models either merge in a failed CE or do not undergo CE at all. Alternative values for an ``optimistic'' CE scenario are given on the lower panel.}
    \label{Tab_rates}
\end{table}

The \textsc{standard} and \textsc{new winds} models display shallow redshift evolution compared to the \cite{Chruslinska19} SFH, but these models more closely recover the LVK merger rate estimate. By contrast, models which use convective penetration (\textsc{new CBM} and \textsc{new winds \& CBM}) models closely trace the SFD while predicting significantly ($\approx3-4$ times) greater BBH merger rates.

We also include estimates for BBH merger rates under an ``optimistic'' CE scenario, in which donors that lack a core-envelope boundary are \emph{not} forced to merge \citep{Belczynski_2008, Dominik_2012}. Such a scenario permits a large population of binaries which interact early in their evolution to merge as BBHs, increasing the merger rate across all models to $\sim 400 \space \rm Gpc^{-3} yr^{-1}$.

\subsection{The BBH Mass and Mass Ratio Spectra}\label{Res_m1}

\begin{figure*}[tb]
    \centering
    \includegraphics[width=1\textwidth]{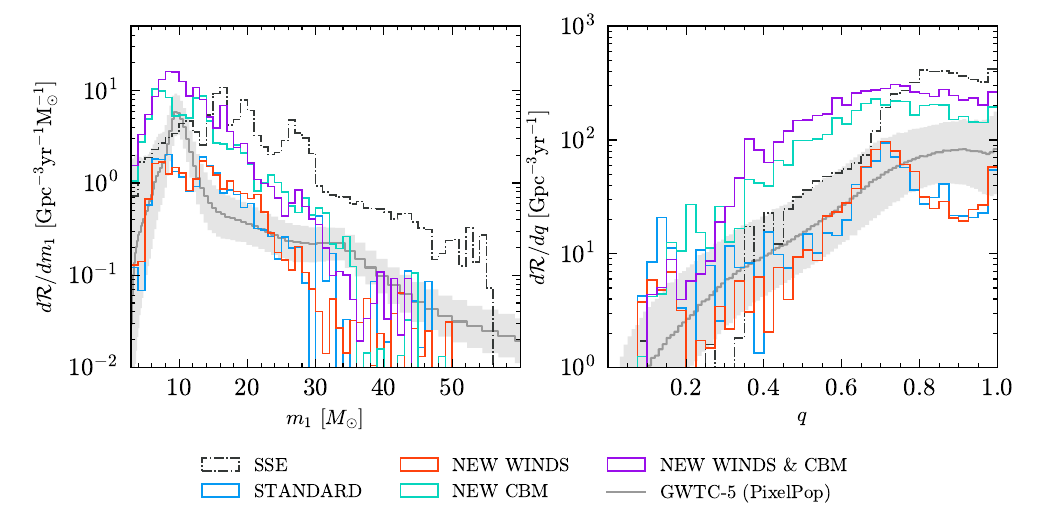}
    \caption{The differential BBH merger rate as a function of $m_1$ (left) and of $q\equiv m_2/m_1$ (right). These distributions include binaries merging at $z<0.2$, as calculated with Equation \ref{Eq_R_obs}.  We include the \texttt{SSE} model (dashed) and the LVK \textsc{PixelPop} posterior (shaded 90\% CI) for reference \citep{ligo2026b_gwtc}. Note the twin features at $m_1\approx 8M_{\odot} $ and $\approx13M_{\odot}$, which approximately flank the LVK peak at $10 M_\odot$. In all \texttt{COSMIC-METISSE} models, the merger rate decreases rapidly above $\approx30 M_\odot$. This decrease is attributable to binary interactions rather than pulsational pair-instability, in agreement with \cite{olejak2026binaryevolutionmimicpairinstability}. By contrast, the \texttt{SSE} model terminates abruptly at $\approx55 M_\odot$ due to pulsational pair-instability.}
    \label{Fig_m1_q_combined}
\end{figure*}


We present the local BBH primary mass spectrum in the left-hand panel of Figure~\ref{Fig_m1_q_combined}. In all \texttt{COSMIC-METISSE} models, the primary mass distribution peaks at $\approx8 M_{\odot}$, and all except \textsc{new winds \& CBM} have a second, significant feature at  $m_1\approx13 M_{\odot}$. While this feature is absent in the \textsc{new winds \& CBM} model, we argue in Section \ref{Disc_SubPops_10Msun} that the same subpopulations are present, but overlap such that the individual peaks are no longer resolved (and have merged to create a single peak at $\approx9M_{\odot}$). Above these ``twin peaks,'' a power law-like continuum extends to $\sim25 M_{\odot}$ (\textsc{standard} and \textsc{new winds}) or $\sim35 M_{\odot}$ (\textsc{new CBM} and \textsc{new winds \& CBM}) before decreasing steeply.

Meanwhile, our reference model obtained with \texttt{SSE} fitting formulae predicts a global maximum at ${m_1\approx16 M_{\odot}}$, with secondary features observable at ${m_1\approx18 M_{\odot}}$ and $25M_{\odot}$. Unlike our \texttt{COSMIC-METISSE} models, the \texttt{SSE} primary mass spectrum terminates abruptly due to pair-instability at $m_1\approx55 M_\odot$.

In the right-hand panel of Figure \ref{Fig_m1_q_combined}, we show the differential local merger rate as a function of the binary mass ratio, $q \equiv m_2/m_1$, with the \textsc{PixelPop} posterior of \cite{ligo2026b_gwtc} included for reference. The mass ratio predicted by \texttt{COSMIC-METISSE} peaks at $q\approx0.7$, but the broadness of the distribution is dependent on the choice of CBM prescription. Models using exponential overshooting (\textsc{standard} and \textsc{new winds}) have prominent peaks at $q\approx0.7$, a feature which is strongly linked to systems that have experienced MRR.  Models using convective penetration (\textsc{new CBM} and \textsc{new winds \& CBM}) peak similarly at $q\gtrsim0.7$, but exhibit a broader distribution. Our \texttt{SSE} model peaks at $q\approx 0.85$ and exhibits an otherwise power law-like structure.

\section{The ``Twin Peaks'' at $10 \space M_\odot$}\label{Disc_SubPops_10Msun}

\begin{figure*}[htbp]
    \centering
    \includegraphics[width=1\textwidth]{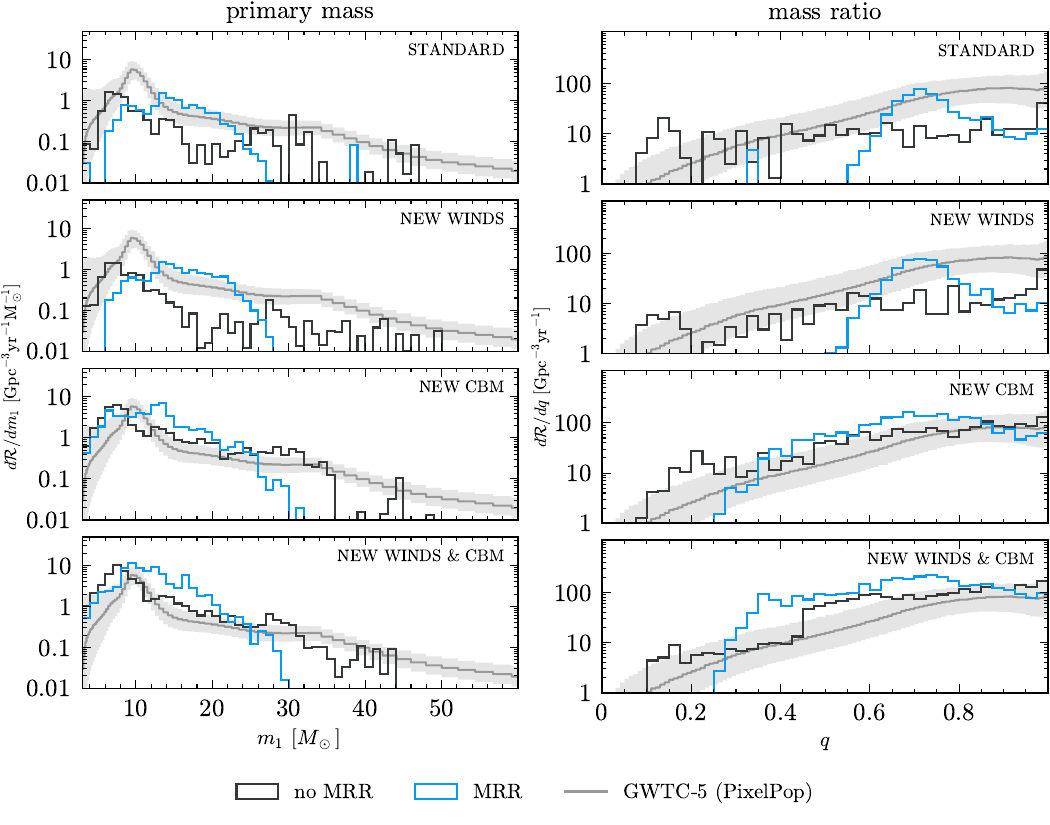}    
    \caption{Differential BBH merger rates as a function of $m_1$ and $q$ for each of our models, split by whether a BBH undergoes MRR. We also show the LVK \textsc{PixelPop} posterior for reference \citep{ligo2026b_gwtc}. BBHs which experience MRR preferentially form with primary masses $>10 M_\odot$, forming the bulk of the MRR ($\approx13 M_\odot$) peak in our results. MRR is also strongly associated with mass ratios $\sim 0.7$, a feature which is especially prominent in the \textsc{standard} and \textsc{new winds} models. Our \textsc{new winds \& CBM} model has only one visible peak formed by the confluence of MRR and non-MRR systems, which are shifted towards $10M_\odot$ by our stellar physics settings.}
    \label{Fig_MRR_hist}
\end{figure*}

\begin{figure*}[tbh]
    \centering
    \includegraphics[width=\textwidth]{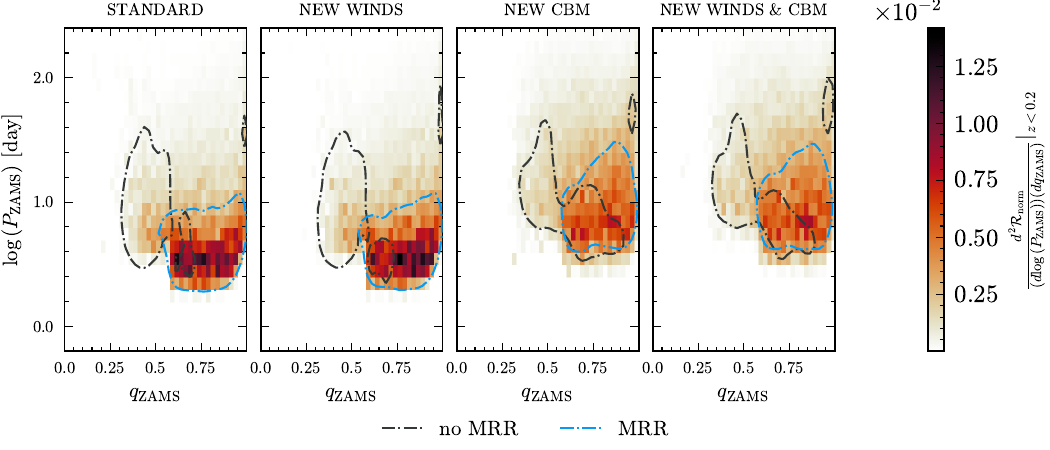}
    \caption{The normalized differential merger rate density parameterized by the ZAMS mass ratio, $q_\text{ZAMS}$,  and the logarithm of the ZAMS period, $\log{(P_{\rm ZAMS})}$. The contour lines denote BBHs which do and do not undergo MRR ($90^\text{th}$ percentile regions). Systems with a natal mass ratio $\gtrsim0.5$ are likely to experience MRR and to form BBHs with $m_1\gtrsim10 M_\odot$. Systems with more unequal mass ratios are unlikely to experience MRR and more likely to form BBHs with $m_1 \lesssim 10 M_\odot$.} 
    \label{Fig_qiPi_2d}
\end{figure*}

As we show in Figure~\ref{Fig_m1_q_combined}, all \texttt{COSMIC-METISSE} model variations except \textsc{new winds \& CBM} produce two peaks located to either side of the inferred LVK maximum at $10M_\odot$. These features are generally placed at $\approx8M_{\odot}$ and $\approx13M_{\odot}$, but their precise locations are somewhat sensitive to our stellar track variations.

These ``twin peaks'' are uniquely associated with the occurrence of mass ratio reversal (MRR), a relationship which we picture in Figure~\ref{Fig_MRR_hist}. BBHs which do not experience MRR preferentially form with $m_1\lessapprox 10 M_\odot$, while those which do not preferentially produce BBHs $\gtrapprox 10 M_\odot$. The outlier \textsc{new winds \& CBM} model, in which we see only one peak at $\approx 9 M_\odot$, still hosts a significant population undergoing MRR, but its location is shifted towards lower masses, effectively overlapping the non-MRR maximum near $10M_\odot$ (see Section \ref{Disc_newWindsCBM}). These MRR BBHs yield a mass ratio distribution which peaks around $q\sim0.7$, while non-MRR BBHs display a continuum-like mass ratio spectrum which climbs towards unity. 

Regardless of the model, the primary determinant of whether a given binary does or does not undergo MRR is the natal mass ratio, $q_\text{ZAMS}$. In Figure~\ref{Fig_qiPi_2d}, we show a 2D histogram of local merger rates parameterized by the logarithm of the initial period, $P_\text{ZAMS}$, and $q_\text{ZAMS}$, as well as contour lines marking the $90^\text{th}$ percentile regions of MRR and non-MRR systems. Non-MRR BBHs are drawn from more extreme birth mass ratios $\lessapprox0.5$, while MRR BBHs evolve from comparatively equal-mass binaries with $q\gtrapprox0.5$. MRR also correlates strongly with lower metallicity, such that the $90^\text{th}$ percentile regions almost completely overlap for $Z\leq 0.5 Z_\odot$. A similar finding was recently presented by \cite{Smith_2026}, but our MRR progenitor systems originate from much lower birth masses ($m_{1,\rm ZAMS} \approx 19-33 M_\odot$, $90^\text{th}$ percentile). Our stellar tracks also influence the range of initial orbital periods from which merging BBHs can evolve. The \textsc{standard} and \textsc{new winds} models draw the bulk of their mergers from $P_\text{ZAMS}\lessapprox5~\rm days$, while models using convective penetration merge binaries with natal periods as high as $\approx40 ~ \rm days$.

To understand these sub-populations, it is prudent that we examine their interactions and formation channels. The vast majority ($\sim 90\%$ across models) of merging BBHs in our data experience at least two mass transfer events. We discuss the first mass transfer, which occurs when the initially more massive star overflows its Roche lobe, in Section~\ref{Disc_SMT_early}. We explore the second mass transfer, between the first-born BH and its evolved stellar companion, in Section~\ref{Disc_CE_late}.

\subsection{The First Mass Transfer is Usually Stable}\label{Disc_SMT_early}

The SMT (or, stable Roche lobe overflow) formation channel has attracted significant interest in recent years \citep{van_Son_2022, briel2026casecaseadetailed, Smith_2026}. Combined with an early MRR, which results from highly conservative mass transfer events, SMT has been invoked as the primary mechanism shaping the BBH merger rate feature at $q\sim0.7$, a feature we identify in all of our \texttt{COSMIC-METISSE} models. We find that the initial mass transfer in merging binaries is generally ($\gtrsim85\%$) dynamically stable.\footnote{The remainder of systems typically undergo unstable mass transfer (CE) before the first supernova; these are predominantly contact binaries with similar component masses, which undergo a double-core common envelope phase.} Due to our adopted thermal timescale-based accretion limit (see Subsection~\ref{Method_COSMIC}), the initial SMT phase can be highly efficient and result in several solar masses being accreted onto the secondary. Most relevantly to the ``twin peaks,'' if MRR occurs in a given system's evolution, it is most likely to happen during this phase.

\begin{figure*}[htbp]

    \centering
    \includegraphics[width=1\textwidth]{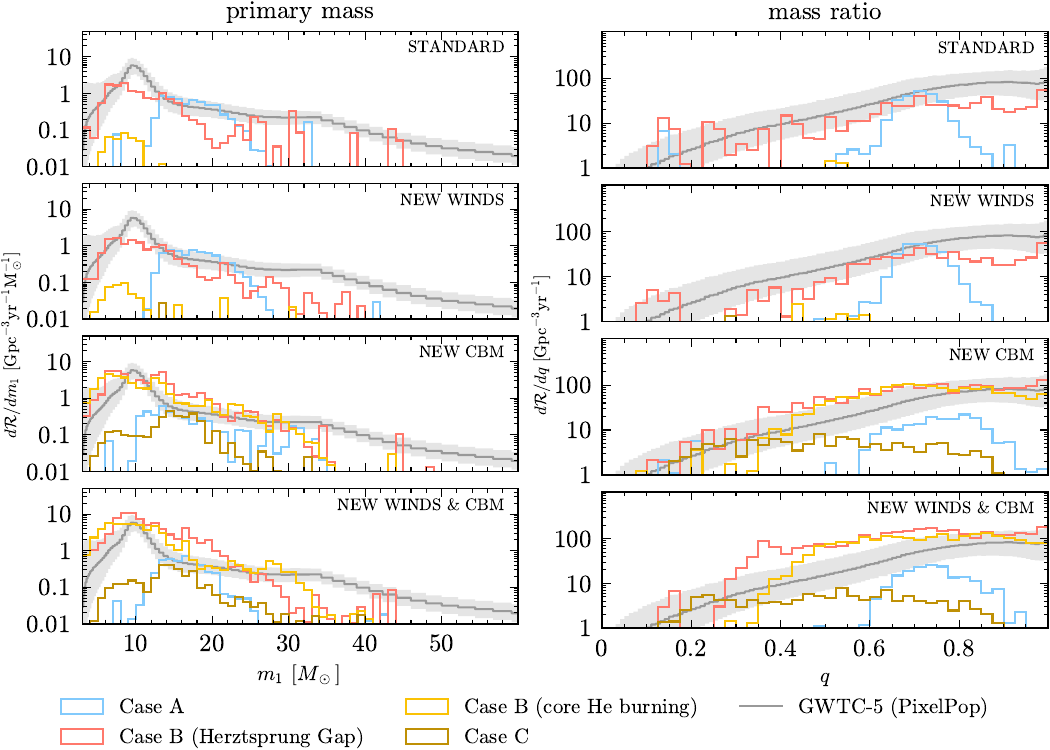}    
    \caption{Differential merger rates as a function of $m_1$ (left set) and of $q$ (right set) colored according to the mass transfer case of the \textit{initial} mass transfer episode. In all models, binaries which undergo Case A mass transfer preferentially form BBHs with primary masses $m_1 \geq 10 M_\odot$ and mass ratios $q\gtrsim0.6$. Binaries which undergo Case B mass transfer produce a prominent peak at $m_1 \approx 8 M_\odot$. In our \textsc{standard} and \textsc{new winds} models, Case A accounts for a majority of mergers with $m_1 \gtrapprox 12 M_\odot$ and forms the bulk of the mass ratio peak at $q\approx0.7$. However, Case A is universally sub-dominant for models using \textsc{new CBM}.}
    \label{Fig_mt_channels}
\end{figure*}

In Figure~\ref{Fig_mt_channels}, we visualize the relationship between BBH primary mass, mass ratio, and the case of the first Roche lobe overflow event in our data. We find a correlation between earlier mass transfer and the MRR sub-population. In the \textsc{standard} and \textsc{new winds} models, MRR BBHs preferentially experience Case A mass transfer, in which the donor star overflows its Roche lobe while on the main-sequence. This is in contrast with non-MRR BBHs, which preferentially experience Case B (post-main sequence) mass transfer. In the \textsc{new CBM} and \textsc{new winds \& CBM} models, Case A is universally subdominant, and a limited contribution from Case C (post-helium burning) donors is present. Even in these models, however, Case A mass transfer is preferentially linked with higher-mass BBHs near the peak of the MRR sub-population.

This evolutionary difference stems from the different radial growth caused by variations in our CBM prescriptions. Convective penetration (\textsc{new CBM}) produces higher main-sequence radii for high mass stars, but their radial growth post-main-sequence is slowed, peaking during core helium burning and post-helium burning phases. By contrast, tracks using only exponential overshooting (\textsc{standard}, \textsc{new winds}) attain their maximum radii shortly after the main-sequence, during the Hertzsprung gap phase (see Figure~\ref{Fig_Rmax}). As a result, binaries evolved using exponential overshooting tend to overflow their Roche lobes earlier.

\begin{figure*}[htbp]
    \centering
    \includegraphics{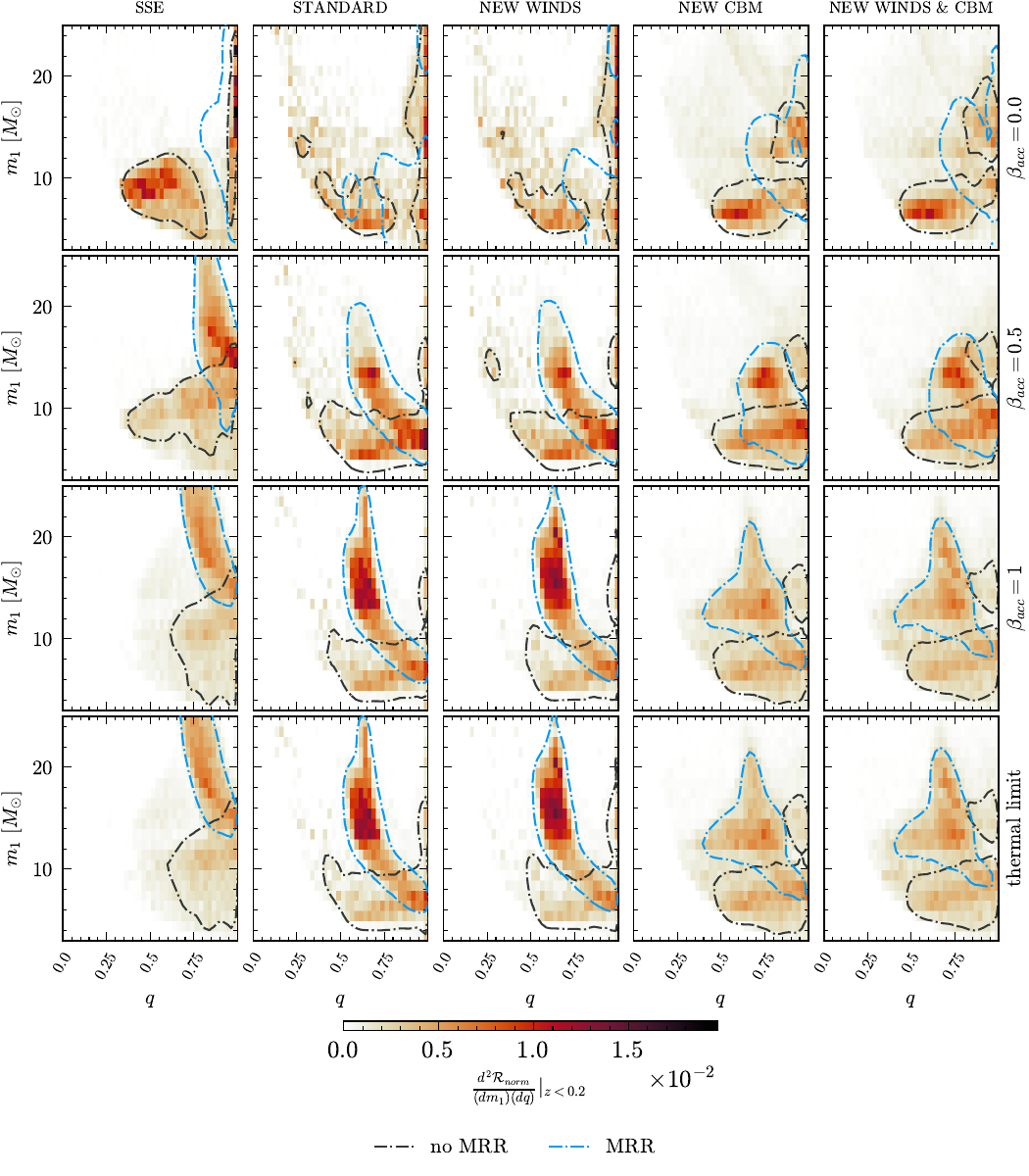}
    \caption{The normalized BBH merger rate density in $m_1$, $q$ space for each of our model variations (columns) and $\beta_\text{acc}$ variations (rows). Our fiducial, thermally-limited models occupy the final row. The 90$^\text{th}$ percentile regions of non-MRR and MRR systems are denoted by dash-dotted lines. These contoured regions reveal two distinct populations flanking the $10M_\odot$ peak. MRR is strongly linked with the higher-mass ($m_1\approx13 M_\odot$) feature which merges with unequal-mass partners ($q\sim0.7$). This sub-population weakens and ultimately disappears when we artificially decrease accretion efficiency by setting $\beta_{\rm acc}\in[0.5, 0]$. Note that an analogous sequence is present in the \texttt{SSE} models, but at greater ($m_1\gtrapprox20$) masses.}
    \label{Fig_beta_comparison}
\end{figure*}

The accretion efficiency of mass transfer events is a key uncertainty in binary stellar evolution \citep{lechien2025binary, Sen_2026}, and it is particularly relevant here given that our models show high rates of MRR. In order to better understand the role of this initial SMT phase, we evolved additional \texttt{COSMIC-METISSE} models with fixed values of $\beta_\text{acc}$ of 0, 0.5, and 1, overriding the thermal timescale-derived limit defined in Equation~\ref{Eq_beta_acc}.

We show the 2D normalized rate density in $m_1$, $q$ for these models in Figure \ref{Fig_beta_comparison}.  For MRR systems, conservative SMT before the first supernova leads to an anti-correlation between the primary mass and the component mass ratio. In the case that this mass transfer event leads to MRR, the resultant BBH is of higher mass and favors unequal mass ratios around $q\sim0.7$. Unsurprisingly, non-conservative mass transfer ($\beta_\text{acc}=0$) essentially eliminates the MRR-linked peak and the feature at $q\sim0.7$. Conservative mass transfer  ($\beta_\text{acc}$ of 0.5 or 1) produces results similar to the thermally-limited case in the $m_1$ and $q$ distributions, and $\beta_\text{acc}=1$ subtly shifts the primary mass distribution upwards towards the MRR peak.

\subsection{The Second Mass Transfer --- Stable or Common Envelope?}\label{Disc_CE_late}

In our samples of merging BBHs, the stability of the second mass transfer event, which occurs between the first-born BH and its stellar companion, is determined chiefly by the CBM prescription in our stellar tracks. BBHs in our \textsc{standard} and \textsc{new winds} models generally undergo only SMT, accounting for $81\%$ and $74\%$ of the merger rate, respectively. BBHs which use convective penetration (\textsc{new CBM} and \textsc{new winds \& CBM}), meanwhile, dominantly undergo a CE phase ($93\%$ and $96\%$, respectively).

Following our assumption that $\alpha_\text{CE} = 1$, the CE phase efficiently exchanges the compact companion's orbital energy to expel the evolved secondary's envelope. This results in hardening of the binary before the second supernova, which permits systems born at much higher initial periods to merge within a Hubble time, as we see in the vertical axis of Figure~\ref{Fig_qiPi_2d}.

The paucity of post-CE mergers in our \textsc{standard} and \textsc{new winds} models compared to those using convective penetration (\textsc{new CBM}) warrants discussion. Models using convective penetration produce a more prominent $10M_\odot$ peak and significantly higher BBH merger rates than those using exponential overshooting alone. Changes to this prescription qualitatively affect binary evolution due to its effect on stellar radius at various life stages. Since our \textsc{standard} and \textsc{new winds} models attain their maximum radii during the Hertzsprung gap, they systematically initiate mass transfer earlier in their evolution (see Figure~\ref{Fig_mt_channels}). The same is true of the second Roche lobe overflow, when the (comparatively more massive) secondary interacts with the (comparatively lighter) first-born BH. A CE is more likely at this stage, and most prospective BBH progenitors in our \textsc{standard} and \textsc{new winds} are forced to merge due to our pessimistic CE assumption. This drastically reduces the proportion of post-CE BBHs in these samples, as well as the overall merger rate.

\subsection{The Birth Metallicity of Merging BBHs}\label{Disc_Metallicity}

\begin{figure}[htb]
    \centering
    \includegraphics[]{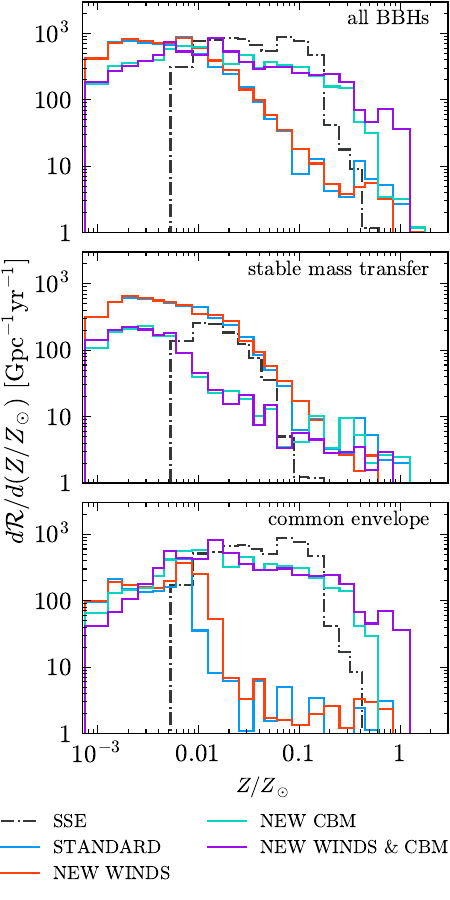}
    \caption{The differential BBH merger rate at $z<0.2$ as a function of metallicity for each of our models, subdivided into binaries which undergo only stable mass transfer (center) and those which undergo at least one CE (bottom). BBHs in stable mass transfer-dominated models (\textsc{standard}, \textsc{new winds}) form primarily from $\le 1\% \space Z_\odot$. The CE-dominated models (\textsc{new CBM}, \textsc{new winds \& CBM}) draw mergers from across the metallicity spectrum. Uniquely, \textsc{new winds \& CBM} contains a large contribution from solar \& super-solar metallicity. Our \texttt{SSE} reference model (black dash-dotted line) is truncated at $Z=0.00014, \space 0.03$ due to the limits of the \cite{Pols_1998, Hurley_2000} formulae.}
    \label{Fig_dRdZ}
\end{figure}

The metallicity distribution of BBH progenitors in our data varies considerably between models, and we visualize these differences in Figure~\ref{Fig_dRdZ}. Since we do not vary the metallicity-dependent star formation history, this variation is purely a result of the BBH formation efficiencies and delay times at each metallicity in our \texttt{COSMIC-METISSE} samples. BBHs in our \textsc{standard} and \textsc{new winds} models form preferentially at very low $Z$ (median $Z\approx 0.02 Z_\odot$), with a steep decrease in merger efficiency between $0.05-1 \space Z_\odot$. In models which use convective penetration (\textsc{new CBM}, \textsc{new winds \& CBM}), merging BBHs form with a higher median metallicity $Z\approx 0.2 Z_\odot$. High metallicities are proportionally well-represented in these models, including solar.

Across model variations, BBHs which undergo only SMT predominantly form at low metallicity, while BBHs which underwent a CE phase display model-dependent behavior. The \textsc{new CBM} and \textsc{new winds \& CBM} variations produces merging BBHs from a broad metallicity continuum peaking at $Z\approx0.2 \space Z_\odot$, in which the bulk of BBHs forming at $> 0.1 \times Z_\odot$ undergo a CE phase. The \textsc{standard} and \textsc{new winds} models efficiently merge BBHs through the SMT and CE channels at $\leq 0.01 \times Z_\odot$, above which both channels decrease in efficiency. The CE channel is subdominant at all metallicities in these models.

Across our models, post-CE binaries tend towards shorter delay times than those which underwent only SMT. Across our \textsc{COSMIC-METISSE} models, the median delay time of post-CE BBHs is $1.78-3.84\,\text{Gyr}$, while SMT BBHs have median delay times of $8.14-8.82\,\text{Gyr}$ (all $50^\text{th}$ percentiles). This helps to explain why our CE-dominated models (\textsc{new CBM} and \textsc{new winds \& CBM}) approximately trace the cosmic SFR, which we show in Figure~\ref{Fig_rates}, while SMT-dominated models do not.

Winds alone have a limited effect on the metallicity spectrum of merging BBHs, shifting the distribution only slightly higher with less powerful \textsc{new winds}. However, the combination of \textsc{new winds \& CBM} is sufficient to generate a substantial merger rate contribution from solar and supersolar ($Z\geq0.014$) metallicity.

\subsection{New Winds \& CBM Bring the Peaks Together}\label{Disc_newWindsCBM}

\begin{figure*}[ht]
    \centering
    \includegraphics[]{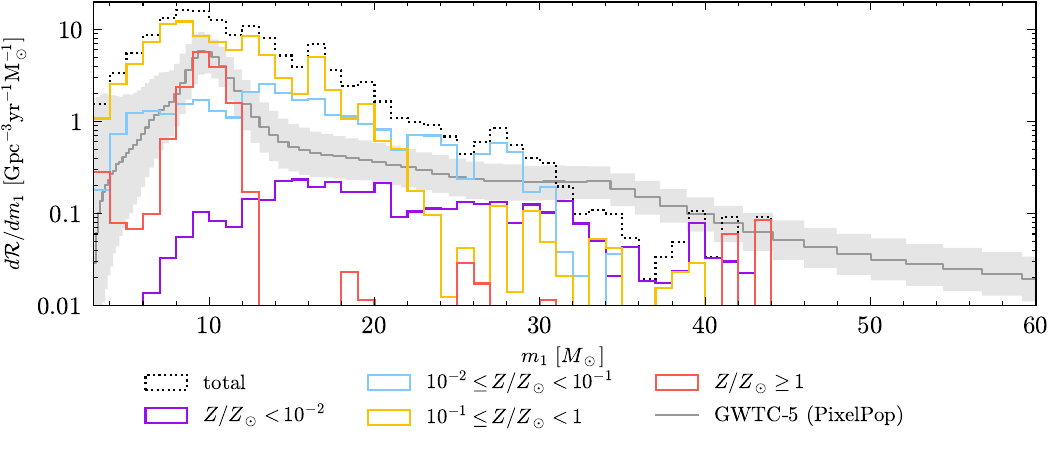}
    \caption{The differential BBH merger rate for our \textsc{new winds \& CBM} model variation, split by birth metallicity. Unsurprisingly, we observe a negative correlation between the metallicity and the primary mass. This model variation uniquely produces a single peak at $\approx8-9 M_\odot$. It also produces a sub-population of high metallicity BBHs at $Z\geq Z_\odot$ ($\geq 0.014$) clustered around the $10M_\odot$ LVK \textsc{PixelPop} maximum \citep{ligo2026b_gwtc}. In order to form this solar metallicity sub-population, both \textsc{new winds} and \textsc{new CBM} are required to preserve the secondary's envelope binding energy and to delay the second Roche lobe overflow, respectively.}
    \label{Fig_dRdm_Z_jdk}
\end{figure*}

Our \textsc{new winds \& CBM} model provides a unique result, as this variation effectively merges the ``twin peaks'' into a single feature near $8-10 M_\odot$ (see Figure~\ref{Fig_m1_q_combined}). This is the most CE-dominant of our models, and closer examination (as in Figure~\ref{Fig_dRdm_Z_jdk}) reveals that this feature is comprised primarily of high-metallicity ($\geq 0.1 Z_\odot$), post-CE binaries. As visualized in Figure~\ref{Fig_dRdm_Z_jdk}, our \textsc{new winds \& CBM} model has strong support at $m_1\approx10 M_\odot$ from BBHs born at solar or supersolar metallicity. This brings this model into closer agreement with the \cite{ligo2026b_gwtc} posterior. This sub-population forms due to the combined effects of our wind and CBM variations. The weak winds of \cite{Krticka2025-rn} and \cite{Decin2023-je}, prevent high-metallicity binaries from widening through winds, while increasing their envelope binding energies. At the same time, the \textsc{new CBM} provided by \cite{Johnston2024-bf}, slows radial growth and delays the second Roche lobe overflow long enough for the binary to survive a CE phase.

We note that while our \textsc{new winds \& CBM} model produces the peak near $m_1\approx 10M_\odot$, it also produces an absolute merger rate $\approx 4.5 \times$ higher than the inferred LVK merger rate \citep{ligo2026b_gwtc}. 

\begin{figure*}[ht]
    \centering
    \includegraphics[]{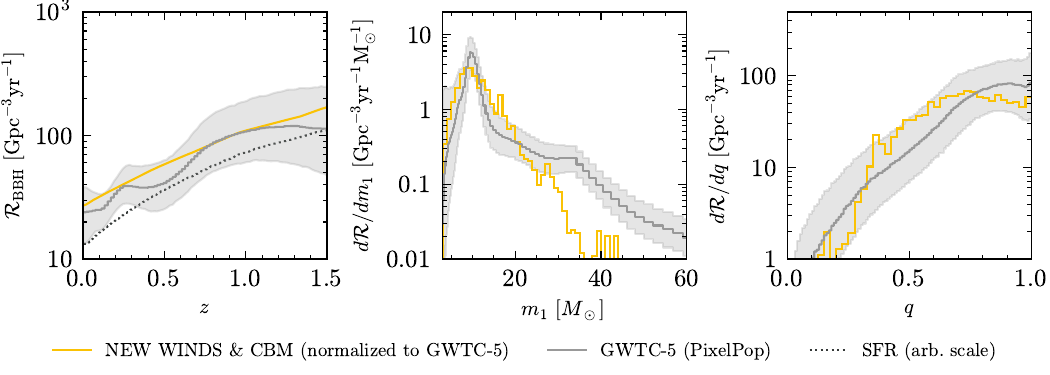}
    \caption{The BBH merger rate (left) primary mass distribution (center), and mass ratio distribution (right) for a hypothetical, normalized \textsc{new winds \& CBM} model. We scaled our model by a factor of $\approxeq4.487^{-1}$, such that the BBH merger rate equals the LVK \textsc{PixelPop} \citep{ligo2026b_gwtc} rate at $z=0.02$. This demonstrative model shows notable agreement in the redshift evolution of the merger rate, the relative height of the $10M_\odot$ feature, and the mass ratio. Future constraints on the redshift-dependent star formation history and the CE formation channel may help models such as ours resolve the formation of merging BBHs through isolated binary evolution.}
    \label{Fig_SFR_scale}
\end{figure*}

For purely demonstrative purposes, we show a hypothetical model in  Figure~\ref{Fig_SFR_scale}, in which we re-normalize the results of our \textsc{new winds \& CBM} model such that its volumetric BBH merger rate equals the \textsc{PixelPop} rate at $z=0.2$ (i.e.~by dividing our merger rates by a factor of $\approx4.5$. This hypothetical, re-normalized model falls within the 90\% CI of the LVK \textsc{PixelPop} posterior, and it shows notable agreement with the inferred LVK primary mass and mass ratio spectra.

As noted by, \cite{Van_Son2023-uc}, the overall shape of the BBH primary mass spectrum is relatively insensitive to uncertainties the metallicity-dependent star formation history. Future studies which constrain the SFH may help to lessen the tension between our models and observational findings.

\section{Discussion \& Conclusions}\label{Conclusions}

Here, we used \texttt{COSMIC-METISSE} to perform rapid population synthesis using interpolated stellar evolutionary tracks whilst varying the input physics for these tracks. Our wind and CBM variations strongly affect the local BBH merger rate and the relative prominence of features in the BBH mass function. We find that the vast majority of merging BBHs undergo at least two mass transfer episodes, of which the first is generally stable. Our \textsc{new winds \& CBM} variation produces a sub-population of high-metallicity binaries at $10M_\odot$, which dominantly undergo MRR. Furthermore, we find evidence across all variations for an anticorrelation between primary mass and mass ratio, with both being associated with MRR. We are unable to replicate this anticorrelation at $\sim10M_\odot$ with the highly-utilized \texttt{SSE} fitting formulae.

We present our key findings in several points:

\begin{itemize}
  \item Our models produce a $\sim10 M_{\odot}$ peak composed of two sub-populations at $\approx 8 M_{\odot}$ and $\approx 13M_{\odot}$. However, the precise location and relative strength of these sub-features is sensitive to details of the stellar physics, namely winds and CBM.  Using newer treatments of both processes merges the two peaks into a single unified peak at $\approx 8-10M_{\odot}$
  \item Membership in these ``twin peaks'' is determined chiefly by the birth mass ratio ($q_\text{ZAMS}$). Progenitor systems with $q_\text{ZAMS} \gtrsim 0.6$ undergo a highly conservative initial SMT episode and preferentially experience MRR. This, followed by a second mass transfer (either stable or unstable, depending on stellar physics) after the first supernova, hardens the binary and leads to the formation of a primary BH with $m_1>10 M_{\odot}$ and $0.6\leq q \leq 0.8$.
  \item Systems with lower initial mass ratios are unable to achieve MRR and form less massive BHs, with $m_1 \leq 10 M_{\odot}$. These BBHs display a continuous mass ratio spectrum that rises towards unity, while MRR BBHs strongly prefer mass ratios $0.6 < q < 0.8$, coinciding with the location of the $q\approx0.7$ knee in the LVK distribution.
  \item While changes to our stellar tracks (winds \& CBM) have a modest effect on the shape of the BBH mass spectrum, they have a profound effect on the overall merger rate. This is primarily accomplished through modulation of the CE formation channel, which is dominant in variations using diffusion-balanced convective penetration (``New CBM'') and sub-dominant in others. When convective penetration is adopted, a CE phase between a BH and its giant partner is more likely to harden the binary enough to merge in a Hubble time. This increases the estimated merger rate by a factor of $\approx4-6$ over models using exponential overshooting alone.
\end{itemize}

\subsection{Caveats \& Future Work}\label{Conclusions_Caveats}

The use of population synthesis codes such as \texttt{COSMIC-METISSE} requires a slew of assumptions that introduce uncertainties in our data products. One of these, the accretion efficiency $\beta_{\rm acc}$, was systematically varied to constrain its effects, while others require further studies to address.

We call particular attention to our adopted $\alpha-\lambda$ CE approximation \citep{Pacznski76}. \texttt{COSMIC-METISSE's} treatment of the CE phase is likely inadequate to fully resolve this rapid and complex process. For example, our ``pessimistic'' CE assumption forces systems which initiate unstable mass transfer with a main-sequence or Hertzsprung gap donor to merge. This is primarily responsible for the low merger rate in our \textsc{standard} and \textsc{new winds} variations. Alternatives to our implementation, such as those that consider the donor's effective temperature \citep{Klencki_2021} or better utilize information about the stars' internal structure \citep{Hirai_2022} may affect our results considerably. Furthermore, \cite{broekgaarden26_ce} found that the fraction of post-CE mergers represented in the local rate is sensitive to various binary evolution parameters, which we do not systematically vary. Further studies are needed to address these uncertainties while taking advantage of detailed single-stellar models.

As mentioned in Subsection~\ref{Disc_newWindsCBM}, the metallicity-dependent cosmic star formation history is a source of considerable uncertainty in our results. Our adopted SFR contains uncertainties of $18\%$ within its extreme cases \citep{Chruslinska19}, and substantial uncertainties remain in how to estimate the SFH as a function of redshift while accounting for low-mass galaxies, starbursts, and observational biases \citep{chruslinska2021impact, chruslinska2024chemical}. 

Additionally, significant developments have occurred in the modeling of supernova explosions, but these are not explored in this study \citep{Burrows_2024, boccioli2024physics}. Our population synthesis pipeline maps evolved stars to remnants through the ``delayed'' prescription of \cite{Fryer_2012}, which is a function of the progenitor's core and envelope mass. Recent, more detailed parameterizations of the core collapse supernova mechanism, which can benefit from detailed stellar models, are available and warrant further investigation \citep{Patton2020, Maltsev2025}.

Finally, we note the similarities between our results and those of \citet{Chen+2026:2026arXiv260626262C} which use the \textsc{SSE} single star tracks via \texttt{COSMIC} to simulate BBH mergers which form exclusively through the SMT channel. While their MRR BBH mergers occur only for primary masses above $20\,\rm{M_{\odot}}$, their MRR mass ratio distribution also peaks near $q\sim0.7$ for conservative accretion during the first mass transfer phase. Similarly, they find that the conditions which define when MRR occurs are dominated by the binary mass ratio at ZAMS and the stability of mass transfer during the first mass transfer phase. Common features which occur in the MRR populations \emph{regardless of single-star assumptions} warrant both more detailed comparisons between population synthesis tools \citep{guerrero2026:inprep} and the application of astrophysically motivated parametric models which utilize these common features \citep{Godfrey+2026:2026arXiv260523083G}.

Our findings show that the BBH merger rate and the relative strength of features in their mass distribution depend sensitively on assumptions about the stellar wind and CBM. \texttt{COSMIC-METISSE} provides an opportunity to test the many assumptions in single-stellar and binary evolution by synthesizing both regimes into a single population synthesis pipeline. Future studies addressing both the above and other uncertainties will help clarify the lives, deaths, and afterlives of BBH progenitors.



\begin{acknowledgments}

We thank Soumendra Roy, Jeff Andrews, Will Farr, and Sylvia Biscoveanu for useful discussions. DM acknowledges support from the National Aeronautics and Space Administration's 2024 Graduate Research Fellowship issued through the North Carolina Space Grant. CR acknowledges support from NASA ATP Grant 80NSSC24K0687 to the University of North Carolina at Chapel Hill, the Research Corporation for Science Advancement (Scialog "Early Science with the LSST", award SA-LSST-2025-123a), a Charles E. Kaufman Foundation New Investigator Research Grant, the Alfred P.~Sloan Foundation, and the David and Lucile Packard Foundation.   DM, KB, MZ, MR, and CR acknowledge support in part from grant NSF PHY-2309135 to the Kavli Institute for Theoretical Physics (KITP), where part of this work was performed.  PA acknowledges support from the European Research Council (ERC) under the Horizon Europe programme (Synergy Grant agreement 101071505: 4D-STAR). KB acknowledges support from the Falco-DeBenedetti Career Development Professorship, \emph{Chandra} Award Numbers GO223031A, GO324037X, and GO425032X, NASA ATP Grant 80NSSC24K0768, and NASA LISA Preparatory Science Program Grant 80NSSC24K0361. AGG was supported by NSF grant PHY-2513312 and by the Simons Collaboration on Black Holes and Strong Gravity through grant SFI-MPS-BH-00012593-07. MZ gratefully acknowledges funding from the Brinson Foundation in support of astrophysics research at the Adler Planetarium.   While partially funded by the European Union, views and opinions expressed are however those of the author only and do not necessarily reflect those of the European Union or the European Research Council. Neither the European Union nor the granting authority can be held responsible for them.
\end{acknowledgments}

\software{
    Astropy \citep{astropy:2013, astropy:2018, astropy:2022},
    cmasher \citep{cmasher},
    COSMIC \citep{Breivik_2020},
    matplotlib \citep{Hunter:2007},
    MC-rates \citep{maclean_2026_20799423},
    METISSE \citep{Agrawal_2020, Agrawal_2025},
    NumPy \citep{harris2020array},
    Pandas \citep{reback2020pandas},
    SciPy \citep{2020SciPy-NMeth},
    xarray \citep{hoyer2017xarray},
    }

\bibliography{
    main,
    mesa,
    mesa_tools,
    software,
    data}{}
\bibliographystyle{aasjournalv7}


\end{document}